\documentclass{aastex}
\usepackage{emulateapj5}
\usepackage{onecolfloatx}

\begin{document}
\twocolumn[

\journalinfo{Accepted by The Astrophysical Journal, astro-ph/0208281}
\title{Probing the Masses of the PSR~J0621+1002 Binary System Through
Relativistic Apsidal Motion} 

\author{Eric M. Splaver and David J. Nice}
\affil{Physics Department, Princeton University\\
Box 708, Princeton, NJ  08544}
\email{esplaver@princeton.edu, dnice@princeton.edu}

\author{Zaven Arzoumanian}
\affil{USRA/NASA Goddard Space Flight Center,
Laboratory for High-Energy Astrophysics\\
Code 662, Greenbelt, MD 20771}
\email{zaven@milkyway.gsfc.nasa.gov}

\author{Fernando Camilo}
\affil{Columbia Astrophysical Laboratory, Columbia University\\
550 West 120th Street, New York, NY 10027}
\email{fernando@astro.columbia.edu}

\author{Andrew G. Lyne}
\affil{University of Manchester, Jodrell Bank Observatory\\ 
Macclesfield, Cheshire, SK11 9DL, United Kingdom}
\email{agl@jb.man.ac.uk}

\and

\author{Ingrid H. Stairs}
\affil{National Radio Astronomy Observatory\\
Box 2, Green Bank, WV 24944}
\email{istairs@nrao.edu}

\submitted{Accepted by The Astrophysical Journal, astro-ph/0208281}

\begin{abstract}
Orbital, spin and astrometric parameters of the millisecond pulsar
PSR~J0621+1002 have been determined through six years of timing
observations at three radio telescopes. The chief result
is a measurement of the rate of periastron advance, 
$\dot{\omega}=0\fdg 0116 \pm 0\fdg 0008$ yr$^{-1}$. Interpreted
as a general relativistic effect, this implies the sum of the
pulsar mass, $m_1$, and the companion mass, $m_2$, to be 
$M=m_1+m_2=2.81\pm 0.30$\,M$_\odot$.
The Keplerian parameters rule out certain combinations of
$m_1$ and $m_2$, as does the non-detection of Shapiro delay
in the pulse arrival times. These constraints, together with 
the assumption that the companion is
a white dwarf, lead to the 68\% confidence maximum likelihood values of 
$m_1=1.70^{+0.32}_{-0.29}$\,M$_\odot$ 
and
$m_2=0.97^{+0.27}_{-0.15}$\,M$_\odot$ and to the 95\% confidence
maximum likelihood values of $m_1=1.70^{+0.59}_{-0.63}$\,M$_\odot$
and $m_2=0.97^{+0.43}_{-0.24}$\,M$_\odot$.
The other major finding is that the pulsar experiences dramatic
variability in its dispersion measure (DM), with gradients as
steep as 0.013 pc\,cm$^{-3}$\,yr$^{-1}$.
A structure function analysis of the DM variations
uncovers spatial fluctuations in the interstellar electron 
density that cannot be fit to a single power law, unlike
the Kolmogorov turbulent spectrum that has been seen in the
direction of other pulsars. Other results from the timing
analysis include the first measurements of the pulsar's proper motion,
$\mu = 3.5\pm 0.3$\,mas\,yr$^{-1}$, and of its spin-down rate, 
$dP/dt=4.7\times10^{-20}$, which, when corrected for kinematic biases 
and combined with the pulse period,
$P=28.8$\,ms, gives a characteristic age of $1.1\times10^{10}$\,yr
and a surface magnetic field strength of $1.2\times10^{9}$\,G. 
\end{abstract}

\keywords{stars: neutron---binaries: general---pulsars: individual 
(PSR~J0621+1002)---relativity---ISM: structure}

]

\begin{table*}[t]
\begin{center}
\caption{Summary of Observations\label{tab:obs}}
\begin{tabular}{ccccccc}
\hline
\hline
\omit\vrule height3pt width0pt\\
Observatory  & Dates           &  Frequency & Bandwidth    & Number of  & Typica
l Integration & RMS residual\tablenotemark{a}\\
             &                 &    (MHz)   &  (MHz)       & TOAs       & Time (
min)          & ($\mu$s)     \\[3pt]
\hline
\omit\vrule height3pt width0pt\\
Arecibo      &  1997.9--2001.5 &  \phn 430  &  \phn 5      &      103   &   29
              &  \phn 2.6   \\
             &  1999.4--2001.5 &  1410      &  10          &  \phn 24   &   29
              &  \phn 3.2   \\[6pt]

Green Bank   &  1995.2--1999.0 &  \phn 370  &  40          &  \phn 48   &   50
              &  17\phd\phn    \\
             &  1995.2--1999.5 &  \phn 575  &  40          &  \phn 49   &   40
              &  13\phd\phn    \\
             &  1995.2--1999.5 &  \phn 800  &  40          &  \phn 39   &   40
              &  19\phd\phn    \\[6pt]

Jodrell Bank &  1996.2--1997.7 &  \phn 410  &  \phn 8      &  \phn 12   &   30
             &  21\phd\phn    \\
             &  1995.7--1999.8 &  \phn 606  &  \phn 8      &      298   &   30
             &  12\phd\phn    \\
             &  1995.7--1997.8 &  1410      &  32          &  \phn 78   &   30
             &  24\phd\phn    \\
             &  1997.8--1999.3 &  1380      &  96          &  \phn 79   &   30
             &  16\phd\phn    \\
             &  1999.7--2001.2 &  1396      &  64          &  \phn 50   &   30
             &  19\phd\phn    \\[3pt]
\hline
\multicolumn{7}{l}{\scriptsize $^a$Values for Arecibo and Green Bank include
the effect of averaging TOAs calculated from shorter integration times.}
\end{tabular}
\end{center}
\end{table*}

\section{Introduction}

Recent pulsar searches have uncovered a new class of binary pulsars.
Most millisecond pulsars are in nearly circular binary
orbits with low-mass He white dwarfs, comprising the
class of so-called low-mass binary pulsars (LMBPs).
The new category of systems, 
the intermediate-mass binary pulsars (IMBPs), have companion stars
with masses above 0.45\,M$_\odot$. Since helium flash occurs at
a core mass of 0.45\,M$_\odot$, the stars must be CO 
or ONeMg white dwarfs.  About eight
IMBPs are currently known. Observationally,
they are distinguished
by high mass functions, $f_1 > 0.015$\,M$_\odot$; by
moderately spun-up pulse periods, 10\,ms $< P <$ 200\,ms;
and by orbital eccentricities that are small, $e < 10^{-2}$, but
somewhat higher than those of the LMBPs \cite{cnst96, clm+01, eb01b}.

This paper reports on timing measurements of PSR~J0621+1002, an IMBP 
with a 28.8-ms spin period in an 8.3-day orbit.
The main goal of our observations was to determine the
pulsar and companion masses through measurement
of post-Newtonian orbital parameters, particularly the
rate of apsidal motion. Measuring apsidal motion
is challenging in white dwarf-pulsar binaries because 
their often extremely small eccentricities---values
of $10^{-5}$ are typical---make it difficult to determine 
the angle of periastron, and hence to measure periastron advance.
A comparatively high eccentricity---still
only $e = 0.0025$---made the detection of apsidal motion 
feasible for PSR~J0621+1002.

\begin{table*}[t]
\caption{Pulse Timing Parameters of PSR~J0621+1002\label{tab:param}\tablenotemark{a}}
\centerline{
\begin{tabular}{ll}
\hline
\hline\omit\vrule height3pt width0pt\\
\multicolumn{2}{c}{Measured Parameters} \\[3pt]
\hline\omit\vrule height3pt width0pt\\
Right ascension, $\alpha$ (J2000)\dotfill       & $06^{\rm h}21^{\rm m}22\fs11108(3)$\\
Declination, $\delta$ (J2000)\dotfill           & $+10\arcdeg02\arcmin38\farcs741(2)$\\
Proper motion in R.A., $\mu_{\alpha}=\dot{\alpha}\cos\delta$ (mas yr$^{-1}$)\dots
                                                & 3.5(3)\\
Proper motion in Dec., $\mu_\delta=\dot{\delta}$ (mas yr$^{-1}$)\dotfill
                                                & $-$0.3(9)\\
Pulse period, ${P}$ (ms)\dotfill                        & 28.853860730049(1)\\
Period derivative, $\dot{P}_{\rm obs}$ ($10^{-20}$)\dotfill
                                                & 4.732(2)\\
Epoch (MJD [TDB])\dotfill                       & 50944.0\\
Orbital period, $P_b$ (days)\dotfill            & 8.3186813(4)\\
Projected semi-major axis, $x$ (lt-s)\dotfill   & 12.0320744(4)\\
Eccentricity, $e$\dotfill                       & 0.00245744(5)\\
Epoch of periastron\tablenotemark{b}, $T_0$ (MJD [TDB])\dotfill
                                                & 50944.75683(4)\\
Longitude of periastron\tablenotemark{b}, $\omega$ (deg)\dotfill
                                                & 188.816(2)\\
Periastron rate of change, $\dot{\omega}$ (deg yr$^{-1}$)\dotfill
                                                & 0.0116(8)\\
Dispersion measure\tablenotemark{c}, DM (pc\,cm$^{-3}$)\dotfill
                                                & 36.6010(6)\\
\hline\omit\vrule height3pt width0pt\\
\multicolumn{2}{c}{Measured Upper Limits} \\[3pt]
\hline\omit\vrule height3pt width0pt\\
Parallax (mas)\dotfill                          & $<2.7$ \\
Pulse period second derivative, $\ddot{P}$ (s$^{-1}$)
                \dotfill                        & $<4\times 10^{-30}$ \\
Orbital period rate of change, $\dot{P_b}$  \dotfill  &  $<5\times 10^{-12}$  \\
Orbital axis rate of change, $\dot{x}$ \dotfill &  $<1.5\times 10^{-14}$   \\
\hline\omit\vrule height3pt width0pt\\
\multicolumn{2}{c}{Derived Parameters} \\[3pt]
\hline\omit\vrule height3pt width0pt\\
Mass function, $f_1$ (M$_\odot$)\dotfill        & 0.027026841(4)\\
Total mass, $M$ (M$_\odot$)\dotfill             & $2.81\pm0.30$\\
Pulsar mass, $m_{1}$ (M$_\odot$)\dotfill        & $1.70^{+0.32}_{-0.29}$\\
Companion mass, $m_{2}$ (M$_\odot$)\dotfill     & $0.97^{+0.27}_{-0.15}$\\
Characteristic age (yr)\dotfill & $1.1\times10^{10}$ \\
Surface magnetic field strength (Gauss)\dotfill & $1.2\times10^{9}$ \\
Total proper motion, $\mu$ (mas yr$^{-1}$)\dotfill  &  $3.5(3)$ \\
\hline\omit\vrule height3pt width0pt\\
\multicolumn{2}{l}{\scriptsize $^a$Values in parentheses are $1\sigma$
uncertainties (68\% confidence) in the last digit quoted.}\\
\multicolumn{2}{l}{\scriptsize $^b\omega$ and $T_0$ are highly covariant. Observers
should use $\omega = 188{\fdg} 815781$ and $T_0 = 50944.756830176$.}\\
\multicolumn{2}{l}{\scriptsize $^c$The DM varies. The value here is the
constant term in an 18-term polynomial expansion
(see \S\ref{subsec:dm}).}\\
\end{tabular}
}
\end{table*}

Mass measurements in white dwarf-pulsar systems can be used
to constrain theories of binary evolution.
The LMBPs and
the IMBPs share roughly similar histories. They both
originate when a giant star transfers mass onto a neutron
star companion, resulting in a spun-up millisecond pulsar
and a white dwarf.
The histories of LMBPs and IMBPs differ, however, in the details
of mass exchange. For LMBPs, it is a stable process that occurs
when the giant swells past its Roche lobe \cite{pk94}.
For IMBPs, it is thought to be an unstable transfer that operates via common
envelope  evolution \cite{vdh94} or super-Eddington accretion
(Taam, King, \& Ritter 2000)\nocite{tkr00}. 
Other pulsar binaries, such as double neutron star systems, evolve
in still other ways.
One way to test binary
evolution scenarios is by comparing the masses of neutron stars in 
different classes of binary systems to infer differences in 
amounts of mass transferred. 
(For a review of pulsar mass measurements, see 
Thorsett \& Chakrabarty 1999\nocite{tc99}.)

The first year of timing observations of PSR~J0621+1002 was
discussed in Camilo et al. (1996)\nocite{cnst96}, which
reported the pulsar's Keplerian orbital elements, position, and 
spin period.  With five additional years of timing data, we
have measured the apsidal motion, spin-down rate,
and proper motion of the pulsar, and we have derived
significant constraints on Shapiro delay. 
We also have found sharp variations in the
dispersion measure (DM), which we use to analyze turbulence in
the interstellar medium (ISM) in the direction of PSR~J0621+1002. 

\section{Observations}

Radio telescopes at Arecibo, Green Bank, and Jodrell Bank recorded 
pulse times of arrival (TOAs) from PSR~J0621+1002 on 529 separate days
between 1995~March~18 and 2001~July~1. Table~\ref{tab:obs} summarizes
the observations. The data comprise (1) three nine-day
campaigns at Arecibo in 1999 May, 2000 May, and 2001 June, each covering
a full pulsar orbit at two radio frequencies; (2) a handful of 
additional Arecibo measurements taken monthly between 1997 and
2001; (3) irregularly spaced observations at Jodrell Bank 
on 387 days between 1995 and 
2001, with an average of five days between epochs; (4) twenty Green Bank 
sessions spaced roughly two months apart, each performed
over four consecutive days at two frequencies; and (5) four 
campaigns covering the full orbit at Green Bank, two each in 1995 and 
1998.

At the 305-m Arecibo telescope, the Princeton Mark~IV data acquisition 
system \cite{sst+00} collected three or four 29-minute data sets
each day at 430 and 1410\,MHz. Local oscillators 
in phase quadrature mixed a 5-MHz passband (10-MHz at 1410\,MHz)
to baseband in both senses of circular polarization. The four
resulting signals were low-pass filtered, sampled, quantized with 4-bit
resolution (2-bit at 1410\,MHz) and stored on disk or tape.  
Upon playback, software coherently dedispersed the voltages
and folded them synchronously at the pulse period over 
190-second integrations, yielding 1024-bin pulse profiles with
four Stokes parameters.

The 140 Foot (43-m) telescope at Green Bank observed the pulsar
for 30 to 60 minutes a day at 370, 575 or 800\,MHz. 
The ``Spectral Processor'', a digital Fourier transform 
spectrometer, divided the signals into 512 spectral channels across
a 40-MHz passband in each of two polarizations. The spectra
were folded synchronously at the pulse period over an integration
time of 300 seconds, producing pulse profiles with 128 phase
bins each. 

The 76-m Lovell telescope at Jodrell Bank carried out a typical observation
of 30 minutes at 410, 606 or 1400\,MHz. The signal was dedispersed
on-line in each of two polarizations using filterbank spectrometers 
with bandwidths of $64\times 0.125$\,MHz for the 400 and 600\,MHz data
and $32\times 1$, $32\times 3$, and $64\times 1$\,MHz for the 1400\,MHz
data. The detected signals were folded synchronously
to make a pulse profile.

In all cases, conventional techniques were used to measure pulse arrival
times. Spectral data were dedispersed and summed to produce a 
single total-intensity profile for a given integration. 
Each profile was cross-correlated with a standard template
to measure the phase offset of the pulse within the profile. 
Different templates were used for each receiver at each
telescope, and for each spectrometer at Jodrell Bank.
The offset was added to the start time and translated to
the middle of the integration to yield a TOA. In a further
step, sets of TOAs from Arecibo and Green Bank were averaged
over intervals of 29 minutes (sometimes longer for Green Bank)
to make a single effective TOA for the interval. Each 
observatory's clock was corrected retroactively to the
UTC timescale, using data from the Global Positioning
System (GPS) satellites.

\section{Timing Model}

\begin{figure*}[t]
\epsscale{1.805}
\plotone{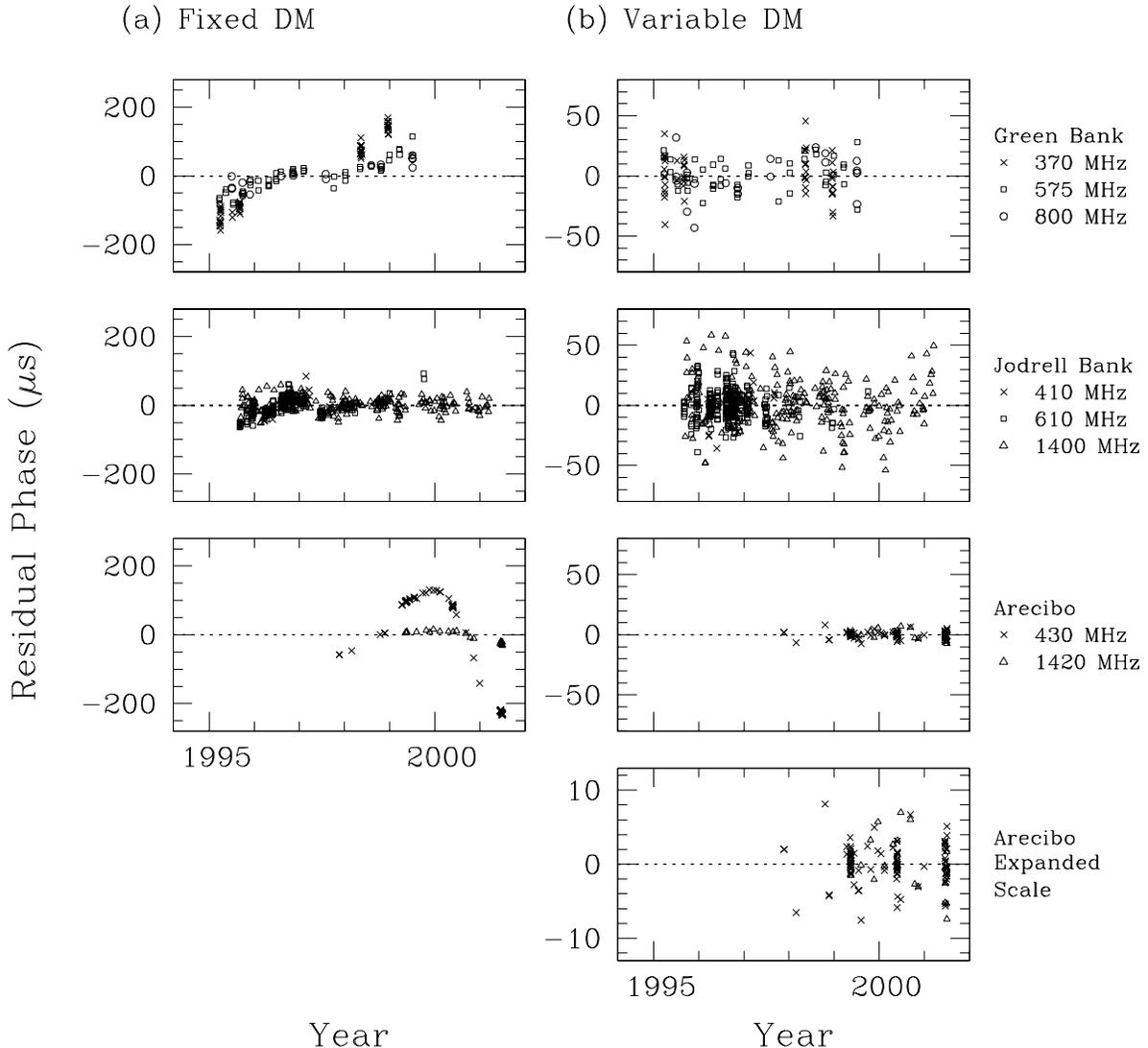}
\caption{Residual pulse arrival times for fits with (a) the DM held
constant and (b) the DM
modeled as an 18-term polynomial. Note the difference in the vertical
scales of the two plots and, in particular, the scale of the
bottom Arecibo panel,
emphasizing the high precision of the TOAs from that observatory.
\label{fig:res}}
\end{figure*}

The TOAs were fit to a model of the pulsar's orbital,
astrometric and spin-down behavior using least-squares
methods. The 
{\sc Tempo}\footnote{http://pulsar.princeton.edu/tempo}
software package performed the fit, 
employing the JPL DE200 solar system
ephemeris and the TT(BIPM01) terrestrial time standard of the  
Bureau International des Poids et Mesures. 
Orbital kinematics were incorporated 
by means of the theory-independent model of Damour \& Deruelle
(1986)\nocite{dd86}. Five Keplerian parameters
(orbital period, semi-major axis projected into the
line of sight, eccentricity, angle of periastron,
and time of periastron passage) and one post-Keplerian
parameter (rate of periastron advance) were necessary
to describe the orbit. Also included in the timing 
model were spin period
and its time derivative, position and proper motion (right
ascension and declination and their time derivatives)
and a time-varying DM (see \S\ref{subsec:dm}).
Besides these astrophysical quantities,
the fit also allowed for arbitrary time offsets
between the TOA sets
to account for possible 
alignment discrepancies between standard templates as well
as for differing signal delays through the various
observing hardware. Table~\ref{tab:param} lists
the best-fit timing parameters.  

Included in the table are upper limits on 
parameters that were not detected. These limits were
found by allowing the extra parameters to vary one at a
time in the fit. In addition to the items in the
table, we carried out an extensive search
for Shapiro delay; see \S\ref{sec:shap}. 

\pagebreak

\subsection{TOA Uncertainties}

Preliminary uncertainties of the TOAs were
calculated in the profile cross-correlations and,
for TOAs made by averaging over many short observations, from the
scatter in the measurements made from the individual short 
observations.  The $\chi^2$ of
the best fit using these uncertainties was high, 
$\chi^2 / \nu = 3.4$, where $\nu = 740$ is the number
of degrees of freedom. Most likely this is because the TOA
uncertainties had been systematically underestimated,
a common problem of unknown origin in high-precision
pulsar timing. To compensate, we added a fixed amount 
in quadrature to the statistical uncertainties
of TOAs in each data set. The amounts were chosen so that
$\chi^2 / \nu \simeq 1$ for each data set in the final
fit. 

\newpage

\subsection{Dispersion Measure Variations} \label{subsec:dm}

Due to dispersion within the ISM, a radio pulse is delayed
in reaching Earth by a number of seconds equal to 
${\rm{DM}}\,/ (2.41 \times 10^{-4}\,f^2)$, where $f$ is the
observing frequency in MHz and the dispersion measure, DM, 
is the column density
of free electrons integrated along the line of sight in
units of pc\,cm$^{-3}$,
\begin{equation}
{\rm{DM}} = \int_{0}^{L} n_e(z)\,dz ,
\end{equation}
in which $L$ is the distance to the pulsar. 
For many pulsars, the DM can accurately be characterized
as a single number that holds steady over years of 
observation. This is not true for PSR~J0621+1002. Figure~\ref{fig:res}a
shows the residual pulse arrival times after removing a model
with a fixed DM. 
Temporal variations in the DM are visible as
secular trends in the residuals. As expected for an effect
that scales as $1/f^2$, the lowest frequencies in the 
figure have the largest residuals. We find the
pulse profile to be stable across the duration of the
observations, so that no part of the trends in 
Figure~\ref{fig:res}a results from intrinsic changes
in the pulsar emission pattern. 

We incorporated DM variations into the timing model using
an 18-term polynomial spanning the entire data set,
\begin{equation}
{\rm DM}(t) = {\rm DM}_0 + \sum_{i=1}^{17} {\rm DM}_{i}(t - t_0)^{i},
\end{equation}
where $t_0$ is the epoch of the parameter fit and the
constant term DM$_0$ is the value quoted for DM in 
Table~\ref{tab:param}. The polynomial coefficients
were simultaneously fit with all other parameters in
the global timing solution.
The ${\rm DM}_0$ term was derived, in effect, from the 575 and
800\,MHz data sets from Green Bank, which were collected
with the same data acquisition system and timed using
the same standard template. The remaining terms 
depended on all the data sets.

Figure~\ref{fig:res}b shows residual arrival times
after subtracting the polynomial DM. These residuals
are consistent with Gaussian noise except perhaps for an
upward rise in the Jodrell Bank points in early 2001, a period
over which all TOAs belong to a single
frequency (1400\,MHz), so that the DM at those epochs
is poorly constrained.  
These same TOAs have some of the largest uncertainties
in the data set, however, and so the unmodeled trend in
them has a negligible effect on the overall timing parameters.  

\subsection{Pulsar Astrometry and Spin-down Behavior}

Using the Taylor \& Cordes (1993)\nocite{tc93} model of the
interstellar free-electron distribution, we estimate the
pulsar distance to be $d = 1.9$\,kpc. A newer model
(J. M. Cordes 2002, private communication) puts the pulsar at a
distance of $d = 1.35$\,kpc.
The spread of distance estimates,
combined with the total proper motion, 
$\mu = 3.5$\,mas yr$^{-1}$, 
gives a relative Sun-pulsar transverse velocity in the range
$V = \mu d = 23\!-\!32$\,km/s. The pulsar is at a small
distance from the Galactic plane,
$z = |d \sin{b}|\lesssim 0.07$\,kpc,
where $b = -2\fdg 0$ is
the pulsar's Galactic latitude.

Camilo et al. (2001)\nocite{clm+01} have noted that IMBPs tend to have a
Galactic scaleheight smaller than that of the LMBPs by a factor of 2--4, 
and that this could
be due to their possessing a space velocity $V$ smaller than that of the LMBPs
by a factor $\la 2$.
The low space velocity of PSR~J0621+1002 supports
this notion. With this number there are now 4 IMBPs with measured
proper motions, for which the space velocities are all approximately
$40 \pm 10$\,km\,s$^{-1}$ \cite{tsb+99,kxc+99}.
These velocities compare
to $V \approx 100$\,km\,s$^{-1}$ for a larger population of millisecond
pulsars 
composed largely of LMBPs and isolated pulsars (Nice \& Taylor 1995; 
Cordes \& Chernoff 1997; Lyne \emph{et al.} 1998; Toscano
\emph{et al.} 1999).\nocite{nt95,cc97,lml+98,tsb+99}. 
A partial explanation for this discrepancy resides in the
different evolutionary histories of the two binary classes: 
LMBP progenitors are $1 + 1.3$\,M$_\odot$ binaries, while IMBPs
may evolve from $4 + 1.3$\,M$_\odot$ systems.
For identical center-of-mass
impulses following the supernova explosion, the pulsars in the latter 
systems will
acquire smaller space velocities.

The observed time derivative of the pulsar's spin period,
$\dot{P}_{\rm{obs}}$, is biased away from
its intrinsic value, $\dot{P}_{\rm{int}}$, as a result
of Doppler accelerations. 
Following Damour \& Taylor (1991)\nocite{dt91}, we find
slight biases due to differential rotation in the plane of 
the Galaxy and due to proper motion. A potential third source of
bias, from acceleration perpendicular to the plane of the Galaxy,
is negligible because of the small distance from PSR~J0621+1002 to the
plane.
With the bias subtracted off, we estimate 
$\dot{P}_{\rm{int}} = (4.3\!-\!4.4)\times 10^{-20}$, about 10\% smaller
than $\dot{P}_{\rm{obs}}$.
 
Under conventional assumptions about the pulsar spin-down
mechanism, the period and intrinsic period derivative yield
a characteristic age of
$\tau = 1.1 \times 10^{10}$\,yr
and a surface magnetic
field strength of
$B_0 = 1.2 \times 10^9$\,G.   

\section{Pulsar and Companion Masses} \label{sec:mass}

\begin{figure*}[t]
\epsscale{2.25}
\plottwo{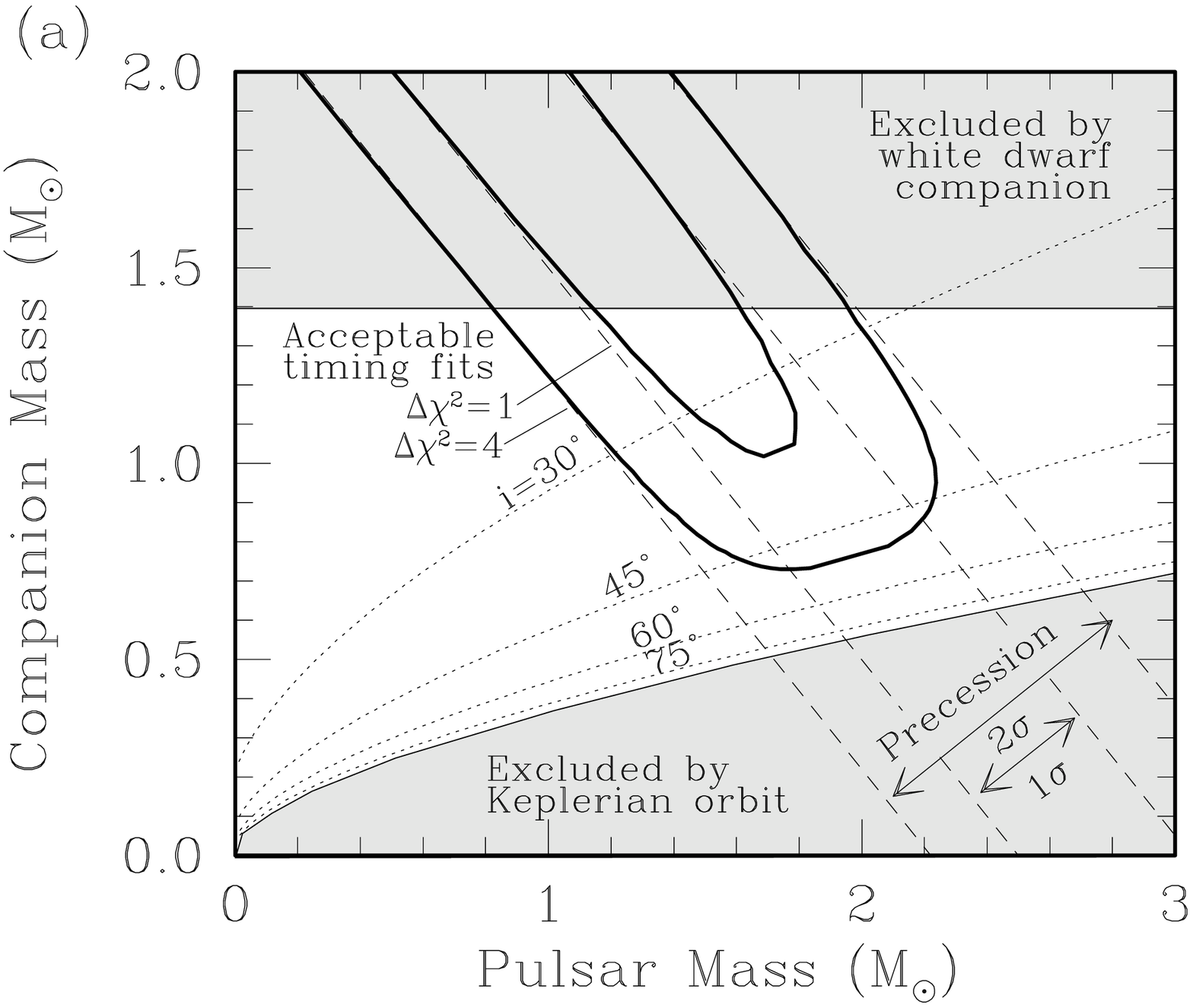}{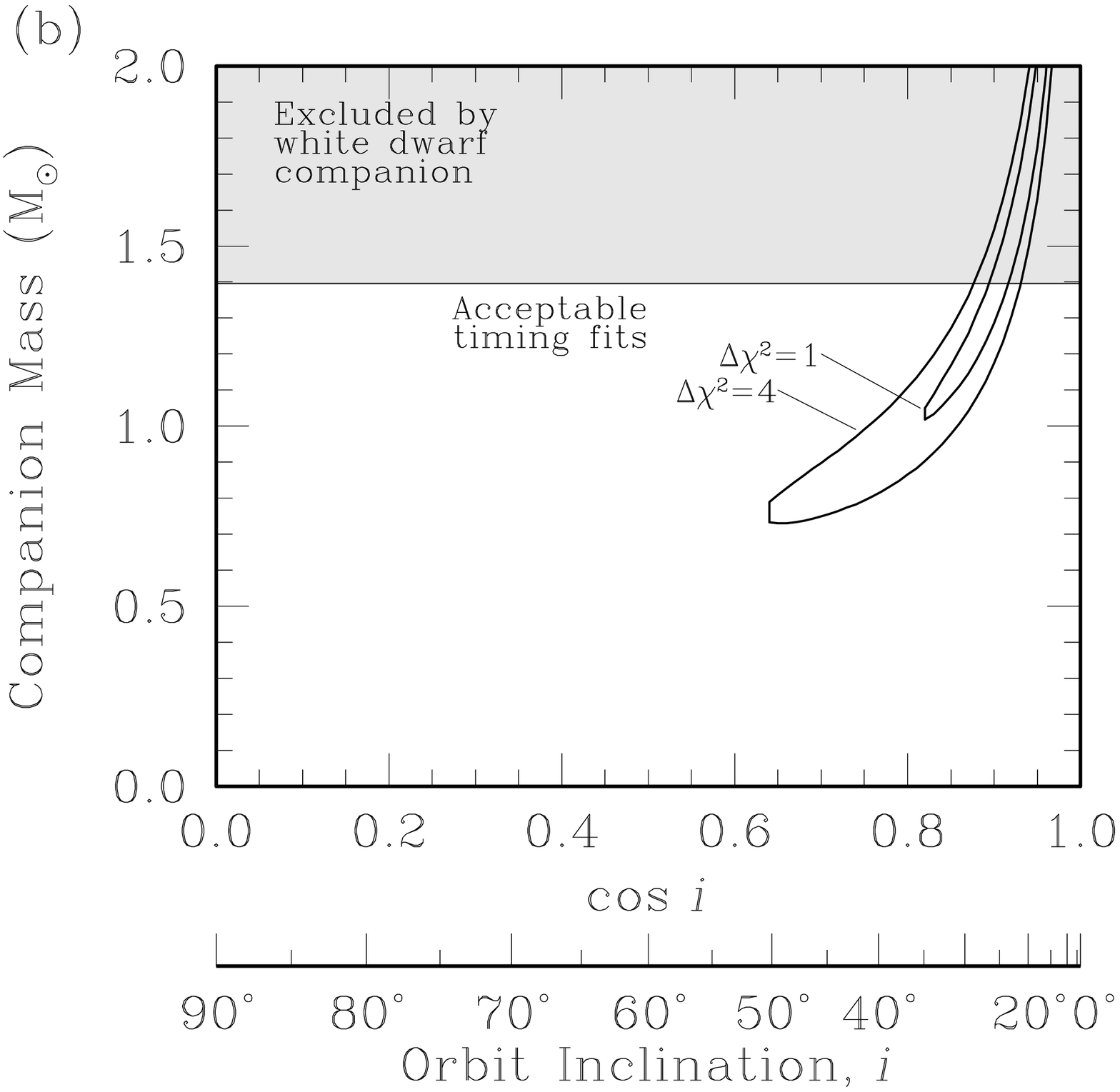}
\caption{Values of pulsar mass, $m_1$,
companion mass, $m_2$, and orbital inclination
angle, $\cos i$, permitted by the Keplerian orbital elements,
the relativistic timing model, and the assumption that the
companion is a white dwarf. (a) Allowed values of $m_1$ and $m_2$.
Dotted lines show selected values of constant $i$.
(b) Allowed values of $\cos i$ and $m_2$.  In both plots,
the true values of the parameters must lie within
the $\Delta\chi^2$ contours and outside the gray regions.
\label{fig:joint}}
\end{figure*}

Our goal is to determine the pulsar mass, $m_1$,
and the companion mass, $m_2$.
The allowed values of the masses are constrained by
the Keplerian orbital elements, the nature of the companion star,
the apsidal motion of the binary system, and the lack of detectable
Shapiro delay. 
In this section, we discuss each of these factors in turn and
display the resulting constraints on the masses in Figure~\ref{fig:joint}a.
The related restrictions on $m_2$ and orbital inclination angle $i$
are shown in Figure~\ref{fig:joint}b. 

\subsection{Keplerian Orbit}

The masses and the inclination angle are related through
the Keplerian mass function,
\begin{eqnarray}
f_1(m_1,m_2,i) \equiv \frac{(m_2 \sin{i})^3}{(m_1 + m_2)^2}
= x^3 \left(\frac{2 \pi}{P_b}\right)^2 \left(\frac{1}{T_\odot}\right) 
\nonumber\\
= 0.027026841 \pm 0.000000004 {\rm \,M}_\odot,
\label{eq:massfunc}
\end{eqnarray}
where
$i$ is the a priori unknown orbital
inclination angle; $P_b$ is the orbital period;
$x \equiv (a_1 \sin{i}) / c$ is the
projected semi-major axis of the pulsar 
measured in light seconds, with $a_1$ the
semi-major axis and $c$ the speed of light; and
$T_\odot \equiv G M_{\odot} / c^3 = 4.925 \times 10^{-6}$\,s,
with $G$ Newton's gravitational constant. 
Since $\sin i\le 1$, equation~\ref{eq:massfunc} can be rewritten to give
an upper limit on $m_1$ in terms of $m_2$,
\begin{equation}
m_1\le m_2^{3/2} f_1^{-1/2} - m_2.
\end{equation}
This constraint is shown as the lower shaded region in
Figure~\ref{fig:joint}a.

\subsection{Upper Limit on Companion Mass} \label{subsec:m2lim}

The companion to the pulsar must be either a main sequence star, a neutron
star, a black hole, or a white dwarf. The first possibility can be ruled out,
as optical observations using the Hubble Space Telescope find
no evidence that the secondary is a main sequence star 
(Camilo et al., to be published elsewhere).
The second possibility can also be eliminated, since it is
improbable that a double neutron star binary could have survived
two supernova explosions and yet retained the 
small eccentricity of the PSR~J0621+1002 system \cite{py98}. 
A pulsar-black hole binary is also unlikely to have such 
a circular orbit \cite{lppo94}.  
The companion must therefore 
be a white dwarf, and as such, its mass, $m_2$, must be
lower than the Chandrasekhar limit, 1.4\,M$_\odot$. This constraint 
is illustrated in Figure~\ref{fig:joint} as the upper shaded
regions.  

\subsection{Relativistic Periastron Advance}

The measured rate of periastron advance, $\dot{\omega}$,
provides another relation between $m_1$ and $m_2$.
For the PSR~J0621+1002 system, we assume that nonrelativistic
contributions to apsidal motion are negligibly small 
(see \S\ref{sec:classical}). 
The general relativistic interpretation of $\dot{\omega}$ 
then yields the combined mass of the stars,
\begin{eqnarray}
M \equiv m_1 + m_2 & = & \left(\frac{P_b}{2 \pi}\right)^{5/2}
\left[\frac{(1 - e^2)\,\dot{\omega}}{3}\right]^{3/2}
\left(\frac{1}{T_\odot}\right) \nonumber\\
& = & 2.81\pm 0.30\,{\rm M}_\odot. 
\label{eq:totmass}
\end{eqnarray}
This constrains $m_1$ and $m_2$ to lie within the strips
indicated by dashed lines in Figure~\ref{fig:joint}a.

\subsection{Shapiro Delay} \label{sec:shap}

According to general relativity, a pulse is delayed as it
propagates through the gravitational potential well of the
secondary. For a pulsar in a circular orbit, this so-called 
Shapiro delay is  
\begin{equation}
\Delta t_s = -2\,m_2\,T_\odot \ln[1-\sin{i} \sin(\phi-\phi_0)],
\label{eq:shap}
\end{equation}
where $\phi$ is the orbital phase in radians and $\phi_0$ is
the phase of the ascending node. For small inclination
angles, the variation of $\Delta t_s$ over an orbit is
nearly sinusoidal, and so it is indistinguishable from a minor
increase in the projected orbital size, $x$.
For edge-on orbits, with $i \approx 90^\circ$, the variation
becomes strongly peaked at $\phi-\phi_0 \approx \pi/2$, when
the pulsar is behind the companion. This breaks the
covariance with the Keplerian parameters, allowing
measurement of Shapiro delay, and hence of $m_2$ and $i$.

Shapiro delay is not detected in the PSR~J0621+1002 timing
data. However, the magnitude of the effect is expected to be
around $15\,\mu$s, which would make Shapiro delay easily detectable
if the inclination angle were large.
The fact that the delay is not observed therefore implies
that the orbit is tilted substantially away from an 
edge-on orientation.

To explore the statistical limits that both the detection
of periastron advance and the non-detection of Shapiro delay
place on $i$ and $m_2$ (and hence on $m_1$), we analyzed
the timing data over a grid of values in the ranges
$0 \leq \cos{i} \leq 1$ and $0 \leq m_2 \leq 2.0$\,M$_\odot$.
For each combination of $m_2$ and $\cos{i}$,
we calculated the Shapiro delay parameters and the
rate of periastron advance according to
general relativity. We then performed a timing fit holding those
quantities fixed while allowing all other parameters to vary.
We recorded the resulting value of $\chi^2$ and its
difference from the global minimum on the grid. Small departures
in $\chi^2$ from the minimum signified the most likely 
configurations of $m_2$ and $\cos{i}$.

The solid contours in Figure~\ref{fig:joint}a show the
regions in which acceptable timing solutions were
found. The interpretation of the
contours is straightforward: within the area allowed
by the precession measurement, the
non-detection of Shapiro delay excludes 
solutions with ``edge-on'' orbits, and so 
the strip is truncated at low values of $\cos i$.
Figure~\ref{fig:joint}b shows that the inclination
angle is constrained with a high degree of confidence
to be less than 50$^\circ$.

\begin{figure}[t]
\epsscale{1.1}
\plotone{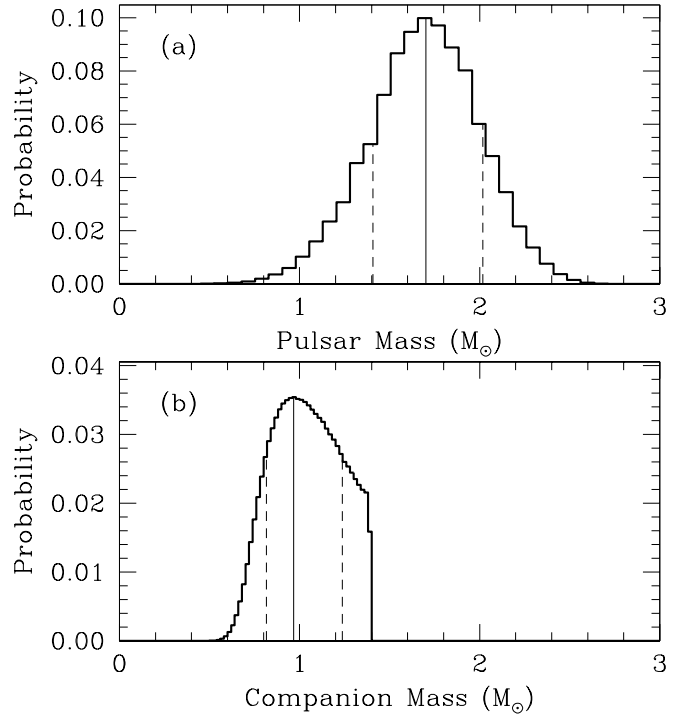}
\epsscale{1}
\caption{Probability distribution functions of (a) the pulsar mass and (b) the
companion mass.
The solid vertical lines mark the maximum likelihood
values of the masses. The dashed vertical lines delimit
68\% confidence regions; they represent the shortest
widths along the mass axes that both enclose the peaks and
contain 68\% of the area under the curves.
For the companion mass, the sharp falloff in the histogram
reflects the assumption that $m_2 \leq 1.40$\,M$_\odot$.
\label{fig:pdfs}}
\end{figure}

Following the statistically rigorous 
procedure described in Appendix~\ref{app:stats}, we
converted the $\chi^2$ differences to probabilities and
derived probability distribution functions (PDFs) for
$m_1$ and $m_2$. The analysis is restricted to timing
solutions for which $m_2 \leq 1.4$\,M$_\odot$, in accordance
with the discussion in \S\ref{subsec:m2lim}. 
The PDFs of the pulsar and companion masses 
appear in Figure~\ref{fig:pdfs}. They yield the maximum
likelihood estimates
\begin{eqnarray}
m_1 & = & 1.70_{-0.29}^{+0.32}\rm{\,M}_\odot 
	\ (\rm{68\% \ confidence}) \nonumber \\
    & = & 1.70_{-0.63}^{+0.59}\rm{\,M}_\odot \ (\rm{95\% \ confidence})
\end{eqnarray}
for the pulsar and
\begin{eqnarray}
m_2 & = & 0.97_{-0.15}^{+0.27}\rm{\,M}_\odot 
	\ (\rm{68\% \ confidence}) \nonumber \\
    & = & 0.97_{-0.24}^{+0.43}\rm{\,M}_\odot \ (\rm{95\% \ confidence})
\end{eqnarray}
for the companion.

Note that the sum of the maximum likelihood
estimates, 2.67\,M$_\odot$, is less than the total system
mass derived in equation~\ref{eq:totmass} from periastron advance alone,
2.81\,M$_\odot$. This is primarily a consequence of the upper
limit on the companion mass, $m_2 \leq 1.4$\,M$_\odot$, 
which preferentially excludes solutions with high total mass, as can
be seen in Figure~\ref{fig:joint}a. 

\ \\

\subsection{Interpretation of the Masses}

A growing body of evidence finds that neutron stars in white dwarf-pulsar
systems are not much more massive than those in double neutron star
binaries, even though the secondaries must lose several tenths
of a solar mass as they evolve toward white dwarfs. In particular, Thorsett \& 
Chakrabarty (1999) \nocite{tc99} find that the masses of neutron stars
orbiting either white dwarfs or other neutron stars  
are consistent with a remarkably narrow
Gaussian distribution, $m = 1.35 \pm 0.04$\,M$_\odot$.
While our measurement of the mass of PSR~J0621+1002
is in statistical agreement with that result, the maximum
likelihood value is suggestively high, allowing the possibility
that a substantial amount of mass was transferred onto the neutron star.

Our estimate of the companion mass implies that the star is 
probably an ONeMg white dwarf and ranks it among the heaviest known white
dwarfs in orbit around a pulsar. As van Kerkwijk \& Kulkarni 
(1999)\nocite{vk99} have pointed out,
there are few massive white dwarfs, $m_2>1\,{\rm M}_\odot$, known
to have evolved from a single massive progenitor star.  They argue
that the companion to PSR~B2303+46, a young pulsar in an eccentric
orbit, is such a white dwarf. Our timing of PSR~J0621+1002 shows
that its companion is similarly likely to have descended from
a massive star, even though the histories of mass loss and
accretion in the eccentric PSR~B2303+46 binary, with no spin-up of the
pulsar, and the circular PSR~J0621+1002 binary, with significant
spin-up, must have been very different.

Given the mass estimates, it seems likely that the 
PSR~J0621+1002 
system formed through a common envelope
and spiral-in phase (Taam, King, \& Ritter 2000;
Tauris, van den Heuvel, \& Savonije 2000)\nocite{tkr00,tvs00}. 
In this scenario, the 
companion originally had a mass of $5-7$\,M$_\odot$, and the pulsar,
initially in a wide orbit with a binary period of a few hundred days,
spiraled in to its current orbit of $P_b = 8.3$ under
a drag force arising from its motion through the envelope. 
The same formation mechanism has been put forward for PSR~J1454$-$5846, 
an object quite
similar to PSR~J0621+1002, with a 12.4-day orbital period,
a 45.2-ms spin period, an eccentricity of 0.002, and a companion mass of
$\sim$~1.1\,M$_\odot$ \cite{clm+01}. 

\newpage

\subsection{Classical Periastron Advance}\label{sec:classical}

The above analysis assumed that the observed
apsidal motion can be entirely attributed to relativity. In principle,
however, apsidal motion could also be caused by distortions of the 
secondary star.
Smarr \& Blandford (1976)\nocite{sb76} considered this possibility 
in the context of a
potential white dwarf companion to the Hulse-Taylor binary pulsar,
PSR~B1913+16. Their analysis can also be applied to PSR~J0621+1002.  
They found
tidal deformation of the companion to contribute negligibly to 
$\dot{\omega}$ for PSR~B1913+16. Since the apsidal advance
per binary period due to tidal deformation 
scales as $a^{-5}$, where $a$ is the major axis,
this effect can also be neglected in the wider PSR~J0621+1002 system.  

A potentially more important effect is rotational deformation, 
which becomes significant if the secondary is spinning rapidly.
The precession rate due to rotation is
\begin{equation}
\dot\omega_{\rm rot} 
 =  nQ
  \left(1-{\textstyle\frac{3}{2}}\sin^2\theta+
         \cot i \sin\theta \cos\theta \cos\Phi_0\right)
\end{equation}
\cite{wex98}, in which $n = 2\pi/P_b$ and 
\begin{equation}
Q = \frac{3k_2 R_2^5\Omega^2}{a^2 G m_2 (1-e^2)^2},
\end{equation}
where $k_2$, $R_2$, $\Omega_2$, and $m_2$ are the
structure constant, radius, angular velocity, and mass of
the secondary; $\theta$ is the angle between the angular momentum
vector of the secondary and the angular momentum vector of the
orbit; and $\Phi_0$ is the longitude of the ascending node
in a reference frame defined by the total angular momentum
vector (see Fig.~9 of Wex 1998).  
Neither $\theta$ nor $\Phi_0$
is known, so we must allow for all possible values in
the ranges $0<\theta<\pi$ and $0<\Phi_0<2\pi$.
Substituting PSR~J0621+1002's Keplerian orbital parameters, the formula
can be written in the
notation of Smarr \& Blandford (1976):
\begin{eqnarray}
\dot\omega_{\rm rot}
&  = &  0\fdg000163\,{\rm yr}^{-1}\,
        \left(\frac{M_\odot}{m_1\!+\!m_2}\right)^{2/3} \nonumber \\ 
& \times & \alpha_6 
     \ \left(1-{\textstyle\frac{3}{2}}\sin^2\theta+
               \cot i \sin\theta \cos\theta \cos\Phi_0\right),
\label{eqn:omdotrot}
\end{eqnarray}
where $\alpha=2k_2R_2^5\Omega_2^2/(3Gm_2)$
and $\alpha_6=\alpha/(10^6\,{\rm km^2})$.
To gauge the largest value that $\dot\omega_{\rm rot}$ could
attain, we will use limits on the masses and 
the orbital inclination angle
from the relativistic analysis, 
recognizing that they would need to be modified should
$\dot{\omega}_{\rm rot}$ be found significant. 
Our observations constrain $20^\circ<i<50^\circ$, so that
$1.2<\cot i<2.7$ (see Fig.~\ref{fig:joint}b), for which the
maximum value of the geometric factor in equation~\ref{eqn:omdotrot}
is 1.7, attained at $\theta=40^\circ$ and $\Phi_0=0$.  
Analyzing models of rotating white dwarfs, Smarr \& Blandford (1976)
found that $\alpha_6\lesssim 15$.  Combining these
restrictions with our measured value of total system mass, the
precession due to rotation for PSR~J0621+1002 can be as high as 
$\dot{\omega}_{\rm rot}\lesssim 0\fdg 0021$\,yr$^{-1}$, about
20\% of our observed value.

There is reason to believe that the classical contribution
to the observed $\dot{\omega}$ is smaller than this.
In most cases, a rotational
deformation will induce a change in the
projected semi-major axis of the orbit, $x=a_1\sin i/c$.
Wex (1998) shows the rate of change to be
\begin{equation}
\frac{\dot{x}}{x} = nQ\cot i\sin\theta\cos\theta\sin\Phi_0,
\end{equation}
from which
\begin{equation}
\dot{\omega}_{\rm rot} = 
  \frac{\dot{x}}{x}\,
  \left(
    \frac{1-\frac{3}{2}\sin^2\theta + \cot i \sin\theta\cos\theta\cos\Phi_0}{%
          \cot i \sin\theta\cos\theta\sin\Phi_0}
  \right).
\end{equation}
Thus, our observed upper limit of 
$|\dot{x}/x|<1.2\times 10^{-15}\,$ implies that
$\dot{\omega}_{\rm{rot}}$ is no more than 
$1.2\times 10^{-15}\,{\rm rad}\,{\rm s}^{-1}
 = 7\fdg 1\times 10^{-6}$\,yr$^{-1}$ times a geometric factor.
Unfortunately, certain special combinations of $i$, $\theta$, and $\Phi_0$
will make the geometric factor large.
For this reason,
we cannot definitively exclude the
possibility that rotational precession contributes
to the observed value of $\dot{\omega}$.
For most cases, however,  
the geometric factor will be of order
unity, and so the upper limit on $\dot{\omega}_{\rm rot}$ 
will be substantially
smaller than our observed value of $\dot{\omega}$. 
Because of this, and because there is no reason to expect
the secondary to be rapidly rotating, we have chosen
to ignore $\dot\omega_{\rm rot}$ in our analysis
of the pulsar and companion masses. 

\section{Density Irregularities in the ISM}\label{sec:ism}
\subsection{Temporal DM Variations}

\begin{figure}[t]
\epsscale{0.90}
\plotone{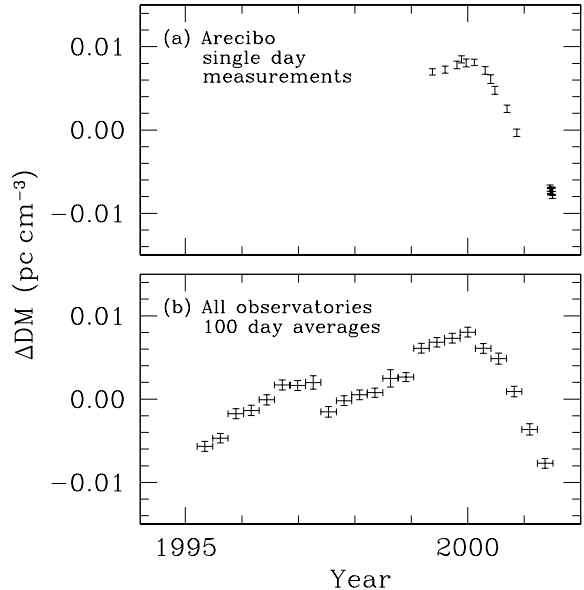}
\epsscale{1.3}
\caption{Dispersion measure as function of time, with
$\Delta\rm{DM} = (\rm{DM} - 36.6010)$.
Vertical bars
are $1\sigma$ uncertainties in the variations.
(a) Measurements within a single day derived from
dual-frequency observations at Arecibo.
(b) Measurements derived from all TOAs averaged
over intervals of 100 days, as indicated by horizontal bars.
\label{fig:dm}}
\end{figure}

Figure~\ref{fig:dm}a shows
the DM of PSR~J0621+1002 calculated
on individual days on which data were collected at Arecibo
at both 430 and 1410\, MHz. A DM drift 
as steep as
0.013 pc\,cm$^{-3}$\,yr$^{-1}$ can be seen in the plot.
If not properly modeled, such a gradient would shift
the TOAs by up to 7\,$\mu$s at 430\,MHz over the 8-day pulsar orbit,
a systematic effect significantly larger than the measurement
uncertainties of the Arecibo TOAs. The DM gradient is among the 
largest ever detected in a pulsar outside 
a nebula \cite{bhvf93}.

Variations in the DM
can be used to investigate
inhomogeneities in the interstellar electron density. 
For this purpose, we modeled
the DM as a series of step functions in time, 
as shown in Figure~\ref{fig:dm}b.
The width of the step intervals, 100 days,
was a compromise between the goal of sampling DM as frequently
as possible and the need for each segment to span enough
multi-frequency data for the DM to be reliably calculated.

In Figure~\ref{fig:dm}b, there appear to be discontinuities in the DM
at 1997.4 and 1999.0. With a careful check of the data around
these dates,   
we confirmed that the DM did indeed change by the amounts
shown within the 100-day time resolution of the figure, and that
the increase and decrease are not artifacts of binning the DM or of
joining together data from different receivers and telescopes.
However, the data do not allow us to distinguish
between nearly instantaneous jumps in DM,
as seen in the Crab pulsar signal
(Backer, Wong, \& Valanju 2000)\nocite{bwv00},
and slower changes on a scale of 100 days.
Given
the rapid but apparently smooth variations in DM seen 
after 2000, we suspect the DM
to be strongly but continuously varying at the earlier epochs as well.

\subsection{Structure Function Analysis}

The DM variations in Figure~\ref{fig:dm} result from the passage
across the line of sight of density irregularities in the ionized ISM.  
The spatial structure of the irregularities can be discerned 
using the two-point structure function of the DM,
\begin{equation}
D_{\mathrm{DM}}(\tau) \equiv
  \left<[{\mathrm{DM}}(t + \tau) - {\mathrm{DM}}(t)]^2\right>,
\end{equation}
where $\tau$ is the time lag between DM measurements and
where the angular brackets denote ensemble averaging (e.g., Cordes
et al. 1990)\nocite{cwd+90}. In a simple one-dimensional model where
the line of sight cuts at the relative Sun-pulsar transverse
velocity $V$ across 
a pattern of ISM irregularities that are ``frozen in'' a thin screen,
and where the thin screen is midway between the Sun and the pulsar,
the time lag is related to the spatial size $l$ of the
density inhomogeneities through $l=V\tau/2$.

\begin{figure}[t]
\epsscale{1.12}
\plotone{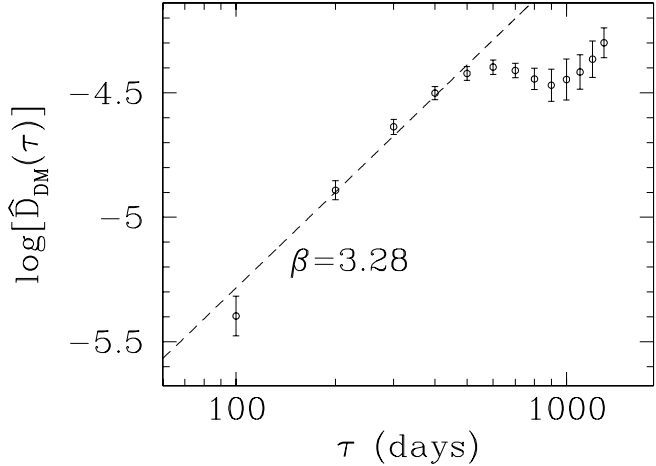}
\epsscale{1}
\caption{Structure function of DM variations
of PSR~J0621+1002
(Fig.~\ref{fig:dm}). The points clearly do not all follow
a single power law. As indicated by the dashed line,
a power law was least-squares fitted to the first five
points according to equation~\ref{eq:taupower}, yielding
the spectral index, $\beta$.
\label{fig:dmstruct}}
\end{figure}

The structure function $D_{\rm DM}(\tau)$ can be approximated from 
the DM values 
at times $t_i$ through the unbiased estimator 
\begin{equation}
\hat{D}_{\mathrm{DM}}(\tau) = \frac{1}{N(\tau)}\sum_{i=1}^{N(\tau)}
  [{\rm{DM}}(t_i + \tau)-{\rm{DM}}(t_i)]^2-\sigma_{\mathrm{DM}}^2,
\label{eq:dmstruct}
\end{equation}
where $N(\tau)$ is the number of DM pairs that enter 
into the summation at a given lag $\tau$ and where $\sigma_{\mathrm{DM}}$
is the mean uncertainty of the DM values.
Figure~\ref{fig:dmstruct} illustrates values of
$\hat{D}_{\mathrm{DM}}(\tau)$ calculated this way for PSR~J0621+1002.
The uncertainties were computed by assuming
Gaussian statistics for the fitted DM values
and formally propagating their covariances
through equation~\ref{eq:dmstruct}. We restrict $\tau$
to values for which $N(\tau) \geq 10$, thereby extending
the lag from a minimum of 100 days to a maximum of
1300 days, about half the length of our full data span. 

\subsection{Not a Simple Power Law}

The structure function
in Figure~\ref{fig:dmstruct} presents a puzzle because
it is not the simple power law that is predicted by  
standard theories of the ISM and that is seen
in the direction of other pulsars. 
In the standard picture, a power law arises from the expectation
that turbulence spreads energy from longer
to shorter length scales. 
Accordingly, the spectrum of perturbations in the electron density
is modeled in terms of the spatial
frequency, $q = 2 \pi / l$, via
\begin{equation}
P(q) \propto q^{-\beta} \label{eq:powerlaw}
\end{equation}
\cite{ric90}. The relation is hypothesized to hold over some
range of wave numbers, $q_i < q < q_o$; below some 
``inner scale'' $q_i = 2 \pi / l_i$ and above some ``outer
scale'' $q_o = 2 \pi / l_o$, various damping mechanisms
are expected to cut off the turbulent energy flow. 
With the assumed linear relation between $l$ and $\tau$, the
structure function becomes 
\begin{equation}
D_{\mathrm{DM}}(\tau) = \left(\frac{\tau}{\tau_0}\right)^{\beta - 2},
\label{eq:taupower}
\end{equation}
where $\tau_0$ is a normalization constant. 
The index $\beta$ is usually predicted to 
be near the Kolmogorov value for turbulence in neutral gases,
$11 / 3$. Studies of DM variations 
in other pulsars have uncovered
power laws with indices close to that value 
(Phillips \& Wolszczan 1992; Kaspi, Taylor, \& Ryba 1994;
Cognard \& Lestrade 1997)\nocite{pw92,ktr94,cl97b}.

The phase structure function in Figure~\ref{fig:dmstruct}
is obviously not a simple power law. It is not clear
how to interpret this. A power law, marked by the dashed
line, can be fit to the first five points in the plot.
The slope of the line yields a spectral index,
$\beta = 3.28 \pm 0.09$, that is reasonably close to 
the Kolmogorov value. At lags longer than 500 days,
the power law fails to hold, although there is a hint
of its reemergence at the longest lags. The flattening
out of the structure function at 500 days is 
a direct consequence of the two sharp breaks in
the DM time series (see Fig.~\ref{fig:dm}b).
If we take the Sun-pulsar
transverse velocity to be $V = 27$\,km\,s$^{-1}$, 
then 500 days 
corresponds to a length of
$l = V \tau / 2= 5.5\times 10^{13}{\rm{\,cm}}$,
implying structure in  the electron density power
spectrum at this scale.

\section{Conclusion}

We have found substantial constraints on the masses of
PSR~J0621+1002 and its orbital companion. The
pulsar mass is found to be $m_1=1.70^{+0.32}_{-0.29}$\,M$_\odot$
(68\% confidence).
The lower end of this uncertainty range is near the canonical
pulsar mass of 1.35\,M$_\odot$, but the mass may also be
several tenths of a solar mass higher, allowing the possibility
that a substantial amount of material accreted onto the
neutron star during the evolution of the system. The mass
of the secondary,
$m_2=0.97^{+0.27}_{-0.15}$\,M$_\odot$ (68\% confidence), 
makes it one of
the heaviest known white dwarfs orbiting a pulsar. 

Can the mass measurements be improved by continued timing observations?
Unfortunately, post-Keplerian effects beyond those considered in this paper,
such as orbital period decay, and gravitational redshift and time dilation,
will not be detectable in the timing data for PSR~J0621+1002 in the
foreseeable future, so any improvement must come about through tighter
measurements of $\dot{\omega}$ and Shapiro delay.

The uncertainty in the measurement of the total mass, 
$M=2.81\pm0.30$\,M$_\odot$, scales linearly with 
the uncertainty of $\dot{\omega}$, which in turn is
inversely proportional to the time span over which data are collected.
This has a simple explanation: the longer the time span of the
observations, the more $\omega$ shifts, and so the easier it is to
measure $\dot{\omega}$. The highest precision data used in this
work---the annual Arecibo campaigns---were collected over two years.
A similar campaign carried out several years in the future would
shrink the uncertainty in $M$ by a factor of a few.

The pulsar and companion masses were further constrained by Shapiro
delay. The precision of the Shapiro delay measurement
(or limit) has no dependence on data span length, so its
uncertainty is reduced only as $n^{-1/2}$, where $n$ is the
number of observations.  At best, a modest improvement could
be made with existing radio telescope resources.

\acknowledgements

The authors wish to thank J. M. Cordes for helpful discussions about
DM variations.
Portions of the Arecibo observations were made as part of the Arecibo
coordinated pulsar timing program; we particularly thank D. Backer, A. Lommen,
D. Lorimer and K. Xilouris for their efforts.  The Arecibo
Observatory is part of the National Astronomy and Ionosphere Center, which is
operated by Cornell University under a cooperative agreement with the National
Science Foundation.  The National Radio Astronomy Observatory is operated by
Associated Universities, Inc., under a cooperative agreement with the National
Science Foundation.  Pulsar research at Princeton University is supported by
National Science Foundation grant 96-18357. FC is supported by 
SAO grant GO1-2063X. IHS is a Jansky Fellow. 

\appendix
\section{Statistical Analysis of Pulsar and Companion Masses}
\label{app:stats}

The PDFs of the pulsar mass, $m_1$, and the companion mass, $m_2$,
shown in \S\ref{sec:shap}
were calculated by least-squares fitting the
TOAs to a timing model across a grid of $m_2$
and $\cos{i}$ values and analyzing the resulting changes in 
$\chi^{2}$. A formal derivation of the PDFs proceeds as
follows. 
If $\chi^{2}_{0}$ is the global minimum
on the grid, then each value of 
\begin{equation}
\Delta \chi^2 (m_2, \, \cos{i}) = \chi^2 (m_2, \, \cos{i}) - \chi^2_{0}
\end{equation}
has a $\chi^2$ distribution with two degrees of freedom.
It therefore maps to a Bayesian likelihood function,
\begin{equation}
p(\{t_j\}\,|\,m_2,\cos{i}) = 
\left(\frac{\textstyle{1}}{\textstyle{2}}\right)\,
e^{-\Delta \chi^2 / 2}, \label{eq:post}
\end{equation}
where $\{t_j\}$ stands for the data set. Accordingly, the joint
posterior probability density of $m_2$ and $\cos{i}$ is 
\begin{equation}
p(m_2, \cos{i}\,|\,\{t_j\}) =
\frac{p(\{t_j\}|\,m_2,\cos{i})}{p(\{t_j\})}\,p(m_2,\cos{i}).
\label{eq:m2cosi}
\end{equation}
The Bayesian ``evidence'', $p(\{t_j\})$, is determined by normalizing
the integral of $p(m_2, \cos{i}\,| \{t_j\})$ over all grid points
that are consistent with $m_1 > 0$, given the relationship among
$m_1$, $m_2$ and $\cos{i}$ in the mass function in 
equation~\ref{eq:massfunc}.
As in all Bayesian investigations, a choice must be made for the prior, 
$p(m_2, \cos{i})$. We selected the product 
of a uniform distribution on $0 \leq m_2 \leq m_{2,\,\mathrm{max}}$
and a uniform distribution on $0 \leq \cos{i} \leq 1$. 
For the companion mass, the choice embodies our ignorance
about the star; all that is 
known for certain is that it is a white dwarf,
for which reason we set $m_{2,\,\mathrm{max}} = 1.4$\,M$_\odot$
(see \S\ref{subsec:m2lim}). For the inclination angle,
a flat distribution in $\cos{i}$ follows from the assumption
that the orbital angular momentum vector has no preferred
direction in space.

The PDF of $m_2$ is obtained by marginalizing equation~\ref{eq:m2cosi}
with respect to $\cos{i}$:
\begin{equation}
p(m_2|\{t_j\}) = \int_{0}^{1} d(\cos{i})\,p(m_2, \cos{i}\,| \{t_j\}).
\end{equation}
Similarly, the PDF of $m_1$ can be expressed as a double integral,
\begin{eqnarray}
p(m_1\, |\{t_j\}) = \int_{0}^{m_{2,\,\mathrm{max}}} dm_2
                  \int_{0}^{1} d(\cos{i})\, \nonumber\\
                  \times p(m_1\, | m_2, \, \cos{i})\,
                  p(m_2, \cos{i}\,| \{t_j\}),
\label{eq:m1pdf}
\end{eqnarray}
where $m_1$ is guaranteed to be consistent with the
mass function by setting 
\begin{equation}
                       p(m_1 | m_2, \cos{i}) =
\delta \left[m_1 - \left(\frac{(m_2 \sin{i})^{3/2}}{f_1^{1/2}} - 
m_2\right)\right],
\end{equation}
with $\delta$ the Dirac function.

\clearpage
\end{document}